# Gate electrostatics and quantum capacitance in ballistic graphene devices

José M. Caridad[1†], Stephen R. Power[2,3,4], Artsem A. Shylau[1], Lene Gammelgaard[1], Antti-Pekka Jauho[1], Peter Bøggild[1]

[1]*Center for Nanostructured Graphene (CNG), Department of Physics, Technical University of Denmark, 2800 Kongens Lyngby, Denmark*

[2]*Catalan Institute of Nanoscience and Nanotechnology (ICN2), CSIC and The Barcelona Institute of Science and Technology, Campus UAB, Bellaterra, 08193 Barcelona, Spain*

[3]*Universitat Autònoma de Barcelona, 08193 Bellaterra (Cerdanyola del Vallès), Spain*

[4]*School of Physics, Trinity College Dublin, Dublin 2, Ireland*

[†]corresponding author: jcar@dtu.dk

**ABSTRACT. We experimentally investigate the charge induction mechanism across gated, narrow, ballistic graphene devices with different degrees of edge disorder. By using magnetoconductance measurements as the probing technique, we demonstrate that devices with large edge disorder exhibit a nearly homogeneous capacitance profile across the device channel, close to the case of an infinitely large graphene sheet. In contrast, devices with lower edge disorder (< 1 nm roughness) are strongly influenced by the fringing electrostatic field at graphene boundaries, in quantitative agreement with theoretical calculations for pristine systems. Specifically, devices with low edge disorder present a large effective capacitance variation across the device channel with a nontrivial, inhomogeneous profile due not only to**

**classical electrostatics but also to quantum mechanical effects. We show that such quantum capacitance contribution, occurring due to the low density of states (DOS) across the device in the presence of an external magnetic field, is considerably altered as a result of the gate electrostatics in the ballistic graphene device. Our conclusions can be extended to any two-dimensional (2D) electronic system confined by a hard-wall potential and are important for understanding the electronic structure and device applications of conducting 2D materials.**

**1. Introduction**

An accurate picture of the electrostatic charge induction mechanism in field effect devices made of graphene or other two-dimensional (2D) crystals [1,2] is necessary to understand the physical properties of these novel nanomaterials [1-18] as well as to guide their exploitation in novel electronic [18], spintronic [19] and optoelectronic [20] applications. Yet, the precise details of the gate electrostatics are far from being understood in these atomically thin and finite-size systems [4-10]. In general, the relation between the gate voltage $V_g$ and the induced average carrier density $n$ is complicated in 2D materials, arising not only from classical electrostatic interactions but also from quantum mechanical effects related to modifications in their band structure under gating [7-10]. The relation $V_g(n)$ is a function of the total capacitance per area of the system $C_{tot}$ and the elementary charge $e$ and can be expressed in terms of classical (electrostatic) $C_c$ and quantum capacitance $C_q$ contributions as [8,9]:

$$V_g = en\left(C_c^{-1} + C_q^{-1}(n)\right) = enC_{tot}^{-1}(n) \quad \text{(Eq.1)},$$

where the dependence of $C_q$ on $n$ makes the capacitance-voltage relationship nonlinear in these systems [8,9]. Thus, $C_{tot}$ is a carrier density dependent property, sensitive to the device geometry, edge morphology and external measurement conditions such as the presence of a magnetic field [9,10]. Furthermore, it exhibits local variations since the charge carrier density $n(\vec{r})$ is not uniform in these finite-size systems: $C_{tot}(\vec{r})/e \equiv \alpha_{tot}(\vec{r}) = n(\vec{r}) \cdot V_g^{-1}$ [5-8]. Here, $\vec{r}$ represents the coordinate(s) in which the carrier density varies and $\alpha_{tot}$ represents the so-called lever arm of the system.

We focus on the capacitance-voltage characteristics of high-quality, ballistic devices made of graphene, the frontrunner of the 2D materials [1,2]. While the injection of charge carriers in graphene field effect devices is often approximated by the infinite parallel-plate capacitor model [1,2] where $n^{\infty} = \alpha_c^{\infty} V_g = (C_{ox}/e) V_g$ and $C_{ox}$ is the gate oxide capacitance per unit area, the presence of hard-wall boundaries leads to accumulation of charges close to the device edges as a consequence of the fringing electrostatic field [7,8] (Fig. 1a). This results first in a generic, inhomogeneous and position dependent effective capacitance profile of classical origin $\alpha_c(x) = C_c(x)/e$ across the device channel ($x$ direction) which qualitatively follows a divergent $1/\sqrt{x}$ dependence towards edges. Quantitatively, $\alpha_c(x)$ will depend on the entire device geometry, with the dominating factor being the ratio $W/b$ between sample width $W$ and the distance to the gate electrode $b$: $\alpha_c(x)$ increases for smaller $W/b$, especially for $W/b \leq 1$ [4, 7-9]. Second, the presence of this varying $\alpha_c(x)$ results in a nontrivial potential profile $U(x)$ across these devices, a fact that may lead to the presence (or even dominance) of a quantum contribution $\alpha_q(x) = C_q(x)/e$ within the total effective capacitance $\alpha_{tot}(x) = C_{tot}(x)/e$ of the system. [7-10]

Such complex gating dependencies are, however, largely unexplored and remain experimentally inconclusive. In particular, the existence of inhomogeneous capacitance profiles of classical origin $\alpha_c(x) = C_c(x)/e$ has been experimentally verified in high-quality, narrow graphene strips up to a certain modulation $\alpha_{tot}(x)^{(max)}/\alpha_c^\infty = C_{tot}(x)^{(max)}/C_{ox}$ [5], where $\alpha_{tot}(x)^{(max)}$ is the maximum measured effective capacitance across the device. However, there are strong discrepancies between different experimental studies [3-6,11-17], and their correspondence with theoretical calculations, too [7-10]. Some studies exhibit capacitance profiles with modulation values across the device $\alpha_{tot}(x)^{(max)}/\alpha_c^\infty \sim 2$, close to those given by classical electrostatic predictions [5]; others report much smaller capacitance modulations ($\alpha_{tot}(x)^{(max)}/\alpha_c^\infty < 1.2$) across similar nanostructures [6], yet others observe no modulation at all ($\alpha(x) = \alpha_c^\infty$) [3], or simply neglect the effect by assuming a constant effective capacitance [11-13]. In a wider perspective, transport studies carried out in high-quality graphene devices display variations in how inhomogeneous gating is accounted for, whether these effects are included [4-6,14,15] or not at all [3,11-13,16,17] during the interpretation of the electrical measurements.

Here, we probe the total effective capacitance profile $\alpha_{tot}(x)$ across ballistic graphene nanoconstrictions (Fig.1b) with different degrees of edge disorder in a perpendicular magnetic field via magnetoconductance measurements. It is noteworthy that the local capacitance is accessible with these measurements in graphene devices [5, 6, 21], similar to conventional semiconductor structures defined in 2D electron gasses [5, 22]. Indeed, magnetoconductance measurements are particularly relevant for the characterization of narrow ($\leq 150$ nm) confined channels [22], where alternative magnetocapacitance techniques are limited due to several reasons. For instance, apart from possessing spatial resolution comparable or larger than the device size, scanning gate microscopy [23], single-electron transistor [24] or microwave impedance spectroscopy [14, 25] measurements might affect or be affected by the actual device electrostatics. Specifically, by analyzing quantum Hall (QH) transport measurements [5,6] we show that devices with stronger edge disorder exhibit a nearly homogeneous effective capacitance profile, with $\alpha_{tot}(x)$ similar to $\alpha_c^{\infty}$. In striking contrast and in quantitative agreement with theoretical predictions for disorder-free systems, devices with a lower degree of edge disorder (< 1 nm roughness) show an inhomogeneous effective capacitance consisting of both classical $\alpha_c(x)$ and quantum $\alpha_q(x)$ contributions. The presence of quantum capacitance effects in graphene devices is a direct consequence of the low density of states (DOS) of the system in a perpendicular magnetic field [10]. Also, $\alpha_q(x)$ is additionally influenced by band-structure modifications occurring in disorder-free, gated graphene devices; systems dominated by the electrostatic screening of the gate potential [4,10]. As shown below, the large $\alpha_c(x)$ diverging toward the edge is comparable in magnitude or larger than $\alpha_q(x)$ at low carrier densities and certain positions $x$ across the channel of these clean devices, so that $\alpha_q(x)$ dominates the total effective capacitance (Eq.1).

**2. Sample fabrication and magnetotransport measurements**

We have fabricated graphene nanoconstrictions (Fig. 1b) with similar lengths $L$ and widths $W$, $L=W \sim 100$ nm on hydrophobized $SiO_2$ substrates with thicknesses $b=100$ nm and with different degrees of edge disorder [4]. Our initial graphene is exfoliated on hydrophobic $SiO_2$, resulting in flakes with mobilities ~ 20000 cm$^2$/Vs [4,26] and mean free paths $l$ larger than $L,W$ at T = 4 K ($l$ ~ 200 nm [4,27] ). Therefore, the transport of carriers through these devices is limited by boundary scattering [4]. Furthermore, we note that the edge roughness in these samples can be precisely assessed via transmission electron microscopy [4], favoring their use in the present work with respect to nanoconstrictions [3] made from encapsulated graphene [28-30].

In particular, we study two types of nanoconstrictions, referred to as type $S_{low}$ and type $S_{high}$ made in an identical way except for the final etching step [4,31]. Sample type $S_{low}$ was etched using reactive ion etching (40 sccm argon, 5 sccm oxygen), a procedure producing significantly less edge disorder than oxygen plasma ashing [4,31] technique used in sample type $S_{high}$. Further fabrication details of these devices can be found in [27] and in [4], where similar systems were studied. Measurements of differential magnetoconductance were performed at different magnetic fields using a Stanford SR830 lock-in amplifier with an excitation voltage of 100 µV at a frequency of 17.77 Hz in a cryostat at T = 300K and T = 4 K.

Figs. 1c,d show the magnetoconductance $G$ as a function of $V_g$ for the two sample types at different perpendicular magnetic fields $B$. The effect of edge disorder reflects itself in the magnitude of the conductance $G(V_g)$, which is more than two times smaller in type $S_{high}$ (Fig. 1c) compared to type $S_{low}$ (Fig. 1d) at any $B$ and $V_g$. Furthermore, sample type $S_{high}$ shows a quantized conductance $G = G(V_g)$ at high $B$ for the zeroth-order Landau level, *LL0*, at the corresponding filling factor for graphene: $\nu = 2$. This plateau is followed by a conductance dip due to the presence of a nonzero longitudinal conductivity in the two-terminal device, effect that depends on the device geometry [4, 32]. For the higher *LL*s, $G$ is smaller than the expected quantized value due to the pronounced effect of edge disorder [4], promoting backscattering between edge channels [33]. In contrast, as reported in our previous study [4], sample type $S_{low}$ (edge roughness < 1 nm) shows a conductance the value of which is larger than the value expected for the single-electron picture and does not exhibit quantization. This effect, referred to as conductance quantization suppression

(CQS) [4,34], is a manifestation of additional, overlapping and counter-propagating conducting channels in the constriction. These channels emerge from a qualitative modification of the bandstructure of the system in the quantum Hall regime due to fringe field effects occurring in ballistic systems with low edge disorder [4,34] (see [27]). As such, their appearance is symptomatic of and relies on the presence of a large and inhomogeneous capacitance profile across the nanoconstrictions [4,34].

## 3. Effective capacitance profile

We experimentally probe $\alpha_{tot}(x)$ in both types of samples $S_{low}$ and $S_{high}$ by using the evolution of the magnetotransport data shown in Figs.1c,d with respect to different $B$ (the selected positions of $G(V_g, B)$ are marked by arrows). This is possible thanks to the extreme sensitivity of the QH transport to the carrier density and carrier density distributions [5,6]. Briefly [5,6], in the QH regime, when the Fermi level lies in-between two $LLs$ and the conduction is governed by $\nu$ propagating edge states, each edge channel probes a different spatial region from the edge of the device [5], whereas the bulk is insulating [5,6]. On the one hand, the carrier density $n_\nu$ for a filling factor $\nu$ is given by $n_\nu = \nu eBh^{-1}$ with a corresponding effective capacitance $\alpha_\nu = n_\nu(V_g - V_{CNP})^{-1}$, where $V_{CNP}$ is the charge neutrality point of the device. On the other hand, an estimate of the position of these edge-channels is given by corresponding cyclotron diameter $d_\nu = (\nu h)^{1/2}(e\pi B)^{-1/2}$ for graphene [5,6]. As the capacitance increase is linked to the charge accumulation across the channel of finite-size graphene devices, edge states of different $LLs$ probe different spatial regions of the device and are subjected to a different effective capacitance. Thus, by making the assignment $\alpha_\nu \equiv \alpha(x = d_\nu)$, the profile of $\alpha_{tot}(x)$ across the device can be experimentally assessed [5,6]. Such capacitance mapping is clear in the absence of back-scattering characteristic of the QH regime.

We argue that this semi-classical method can be also used in our ballistic devices exhibiting the CQS effect (Sample type $S_{\text{low}}$, Fig. 1d). This is particularly valid at gate voltages where the (nonquantized) conductance shows a maximum (arrows in Fig. 1d) $G^{(\text{max})}$, places where, if existent, backscattering between the counter propagating edge channels in the system is weakest [4,34] (see [27]). As such, the link between $G$ and $n(x)$ is still valid by taking $\nu(x) \sim G^{(\text{max})}(e^2/h)$, and the capacitance profile $\alpha_{tot}(x)$ can then be extracted as demonstrated below when comparing the experimental values with the corresponding theoretical calculations.

Fig. 2 shows the extracted effective capacitance $\alpha_\nu$ versus $d_\nu$ plotted for the different *LLs* in our samples with high and low edge disorder. Samples with high edge disorder (sample type $S_{\text{high}}$, Fig. 2a) show a slowly monotonically increasing capacitance $\alpha_\nu(d_\nu)$ towards edges, with quantitative values close to those of the infinite parallel-plate capacitor, $\alpha_c^\infty$ (green dotted line). Specifically, this capacitance profile is nearly homogeneous across the device channel ( $x$ direction), with only a small measured modulation up to $\alpha_{tot}(x)^{(\text{max})}/\alpha_c^\infty \sim 1.2$ for *LL0*. In contrast, devices with low edge disorder (sample type $S_{\text{low}}$, Fig.2b) exhibit an inhomogeneous capacitance profile, with much larger modulations ($\alpha_{tot}(x)^{(\text{max})}/\alpha_c^\infty \sim 2.5$ for the case of *LL0* and $\sim 1.7$ for *LL1*). Moreover, we note how the capacitance profile is nonmonotonic in these samples for *LL0*.

Three initial conclusions can be drawn from the experimental data shown in Fig.2, all of them consistent with a more pronounced capacitance profile for sample type $S_{\text{low}}$ as compared to $S_{\text{high}}$. *(i)* As expected from an electrostatic point of view [6,35], there exists electron-hole ($e^-$ – $h^+$) symmetry in both type of samples. The difference in capacitances observed for e[-] and h[+] is lower than 10% with respect to the absolute capacitance at any $B$ *(ii)* The lowest *LL* (*LL0*) is closest to the edge, where more charge is expected to accumulate [4-8]. Consequently, for a constant $B$, $\alpha_v$ exhibits a higher absolute value for *LL0* than for *LL1* in sample type $S_{low}$. In addition, *(iii)* for a varying $B$, the capacitance variation of *LL0* is the largest as well.

Next, we analyze in more detail the capacitance profile $\alpha_v(d_v)$ in both types of devices. For *LL0*, the small modulation $\alpha_{tot}(x)^{(max)}/\alpha_c^\infty \sim 1.2$ in devices of sample type $S_{high}$ differs quantitatively from the electrostatic capacitance $\alpha_c(x)$ calculated for our constriction geometry, which is monotonic and diverges rapidly towards edges (Fig. 3, blue line; simulation details in [27]). Such behavior is attributed to scattering from edge defects, decreasing the charge accumulation in a similar manner as seen for low quality devices with disorder in the channel [5,6]. This result explains the fact that some high-quality devices in literature exhibit effective capacitances resembling the one from an infinite parallel-plate capacitor $\alpha_c^\infty$ [3,4].

Then, we note that despite having a large modulation, the capacitance variation in sample type $S_{low}$ does not completely follow the trend of the calculated classical electrostatic capacitance $\alpha_c(x)$ either: rather than following a divergent $1/\sqrt{x}$ dependence towards the edge [7,8] the measured capacitance $\alpha_v$ has a nonmonotonic profile, decreasing more rapidly than $\alpha_c(x)$ for distances close to the edge and in the middle of the channel. Moreover, $\alpha_v$ exhibits some values smaller than $\alpha_c^\infty$ at the central part of the constriction (distances > 35 nm from the edge) for *LL0* but not for *LL1*. In addition, $\alpha_v$ shows a maximum $\alpha_v^{(max)}$ at a position $d_{v,max} \sim 24$ nm from the edge, with $\alpha_v$ smaller than $\alpha_v^{(max)}$ for $d_v < d_{v,max}$. We do not expect this extremal value to be caused by the breakdown of the classical capacitance $\alpha_c(x)$ at distances from edges given by the magnetic length $l_B = \hbar^{1/2}(eB)^{-1/2}$ [7]: $d_{\nu,\max}$ is three times larger than $l_B$ at the corresponding $B$ (10 T) at which the maximum is observed. This argument is additionally supported by the fact that no local maximum is experimentally observed in sample type $S_{high}$ (Fig.2a) despite the minimum measured $d_\nu$ in such samples (15 nm) being smaller than $d_{\nu,\max}$.

**4. Effective capacitance calculation: classical and quantum capacitance contributions**

We demonstrate that the mismatch between the calculated electrostatic capacitance (Fig. 3, blue line) and the experimentally probed capacitance of sample type $S_{low}$ at the *LL0* (Fig 2b, black circular dots and Fig 3, inset) can be fully accounted for by including quantum contributions to the total capacitance of the system. Such quantum contributions, arising due to the low local DOS [36-39] across the quasi- one dimensional system [10], are not only dependent on the presence of external magnetic fields $B$ but also heavily affected by the electrostatic screening of the gate potential [4,34] (see [27]).

We calculate the quantum capacitance contribution [27] $\alpha_q(x)$ across the ballistic, gated devices at $B$ = 10T and low carrier densities (*LL0*) (Fig. 3, red line). First, we show that both classical and quantum contributions are comparable in magnitude and coincide at two distances: ~ 25 nm far from the edge and at the central part of the nanodevice (~40 nm far from the edge). By using Eq.1, we calculate the total effective capacitance $\alpha_{tot}(x)$ of these devices (Fig. 3, black curve), which is in quantitative agreement (below ~30% mismatch) with our experimental data $\alpha_\nu(d_\nu)$ for *LL0* (Fig 2b and Fig. 3, inset). Specifically, $\alpha_{tot}(x)$ shows a clear maximum $\alpha_{tot}^{(\max)}(x_1)$ at a position $x_1 \sim 25$ nm, close to $d_{\nu,\max}$. For closer distances ($x < x_1$) towards the edge at $x = 0$, $\alpha_{tot}(x)$ is dominated by quantum contributions since $\alpha_c(x) > \alpha_q(x)$. Instead, for distances $x_1 < x < x_2$ with $x_2 \sim 32$ nm, $\alpha_{tot}(x)$ is mostly dominated by the classical capacitance profile, since $\alpha_c(x) < \alpha_q(x)$ in this interval. Furthermore, for distances $x > x_2$ (central part of the nanoconstriction) $\alpha_{tot}(x)$ is again dominated by the quantum contribution due to the vanishing local DOS at these positons. This argument explains the fact that experimentally, $\alpha_v$ for *LL0* is smaller than $\alpha_c^\infty$ at the central part of the constriction, too. Finally, for completeness, we note that the experimental capacitance profile $\alpha_v$ shown for *LL1* in samples $S_{low}$ (Fig 2b, red triangles) increases monotonically towards edges, similar to the classical profile. This behavior is due to the fact that the quantum contribution to the capacitance for *LL1* is larger than the classical one [27] for the probed distances from the edge $x \equiv d_v$ (between 30 and 40 nm). Thus, the classical contribution dominates the total effective capacitance profile at these higher carrier densities (*LL1*).

## 5. Conclusions

In summary, by analyzing quantum Hall transport measurements, we have extracted the total effective capacitance profile across ballistic graphene devices. Such profiles heavily depend on the edge disorder level in the device. In particular, similar to diffusive devices, ballistic samples with a large edge roughness show a nearly homogeneous capacitance with small modulation and values close to the infinite parallel-plate capacitor. In contrast, in excellent agreement with calculations for pristine and gated systems, narrow ballistic devices with low edge disorder (< 1 nm roughness) show an effective capacitance profile with a much larger modulation due to an enhanced impact of the fringe field effects at graphene edges. This profile is nonmonotonic at low carrier densities due to the interplay between classical and quantum capacitance contributions across the gated device even when using relatively thick (100 nm) dielectric layers [8,9]. Despite being demonstrated in ballistic graphene nanoconstrictions, our conclusions can be extended to other systems with sharp edge potential [4,7,8,34], and thus, our findings can help to understand the electronic properties of other types of gated nanostructures made from 2D materials [40,41] or other 2D systems [42].

**Acknowledgements** We acknowledge stimulating discussions with K. Kaasbjerg. This work was supported by the Danish National Research Foundation Center for Nanostructured Graphene, project DNRF103, and the Union's Horizon 2020 research and innovation programme (grant agreement No GrapheneCore2 785219). J.M.C. acknowledges funding from the Øtto Monsteds Fond. S.R.P. acknowledges funding from the European Union's Horizon 2020 research and innovation programme under the Marie Skłodowska-Curie grant agreement No 665919 and from the Irish Research Council under the laureate awards programme.

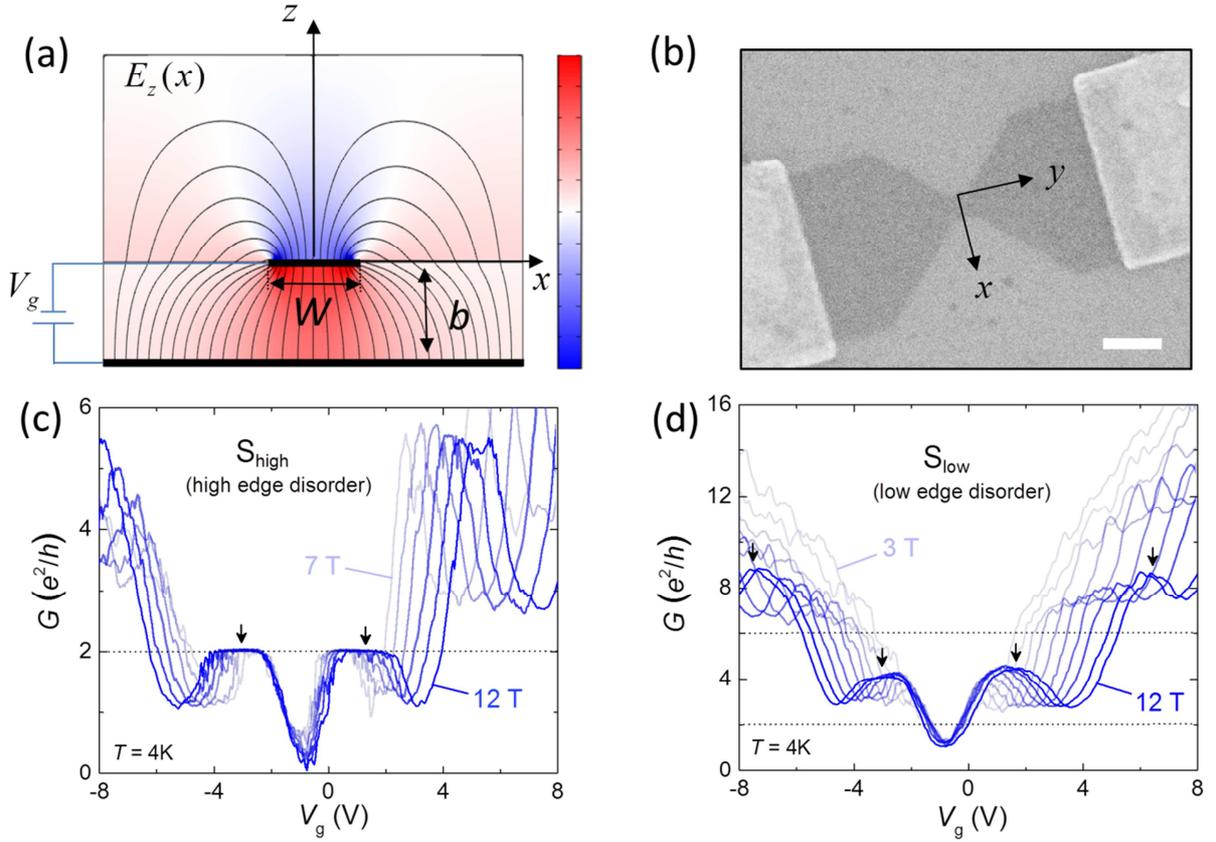

**Fig. 1. Sample fabrication and electrical characterization** (a) Electric field distribution $E_z(x)$ around a gated graphene device with $b=W=100$ obtained by solving the Poisson equation in the device [27]. (b) Scanning electron micrograph of a graphene nanoconstriction device. Scale bar is 200 nm. (c) Conductance $G$ vs gate voltage $V_g$ in a nanoconstriction of type $S_{high}$ at different magnetic fields (from 7 to 12 T in steps of 1T) at $T = 4$ K. (d) Conductance $G$ vs gate voltage $V_g$ in a nanoconstriction of type $S_{low}$ at different magnetic fields (from 3 to 12 T in steps of 1T) at $T = 4$ K. Arrows in panels (c) and (d) indicate the experimental points $(V_g, G)$ taken to calculate $\alpha_\nu(d_\nu)$ in Fig. 2 for B = 12T, i.e. center of the plateau for sample $S_{high}$ [5,6] and $G^{max}$ for $S_{low}$. Corresponding points are taken for other magnetic fields. Dotted lines indicate the expected quantization values for graphene.

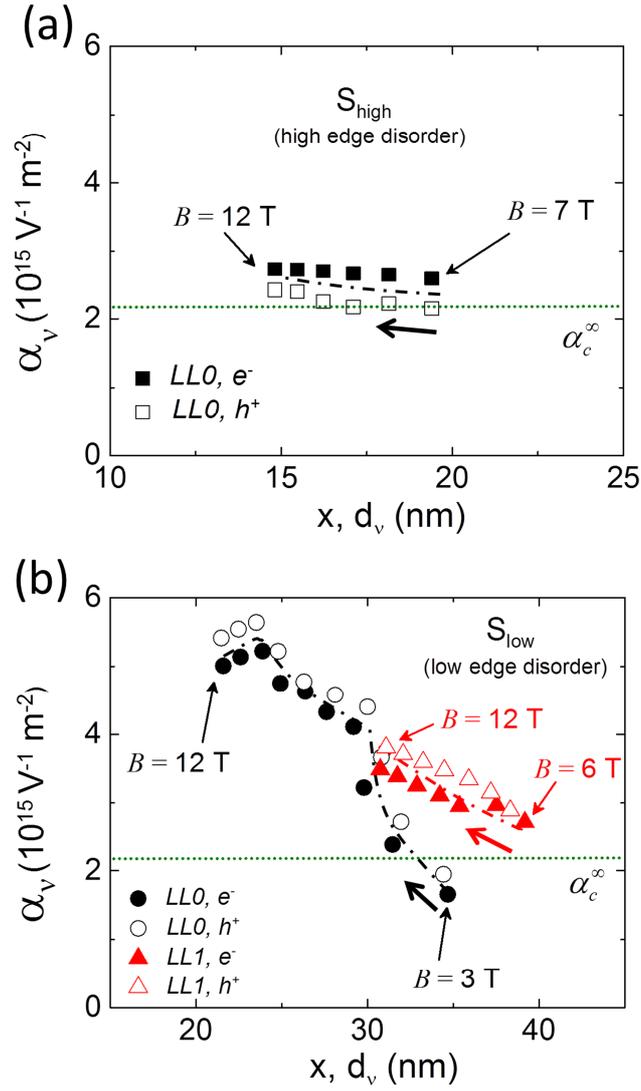

**Fig. 2.** Experimental effective capacitance $\alpha_v$ across the nanoconstrictions for different Landau levels (*LLs*) for electrons ($e^-$) and holes ($h^+$) at positions $x$ corresponding to the cyclotron diameter $d_v$. (a) Sample type $S_{high}$. $B$ varies from 7 to 12 T increasing in 1-T steps. B = 7 T is the minimum $B$ to observe a conductance plateau for *LL0* in samples $S_{high}$. (b) Sample type $S_{low}$. $B$ varies from 3 to 12T (*LL0*) and from $B = 6$ T to B=12T (*LL1*) in 1-T steps. $B = 3$T and $B = 6$T are the minimum fields to observe the CQS regime in these samples for *LL0* and *LL1*,

respectively. Green dotted lines in both panels show the capacitance value for an infinite parallel-plate capacitor $\alpha_c^\infty = \frac{C_{ox}}{e} = 2.19 \times 10^{15} \ V^{-1}m^{-2}$. Dash-dotted lines represent a guide to the eye.

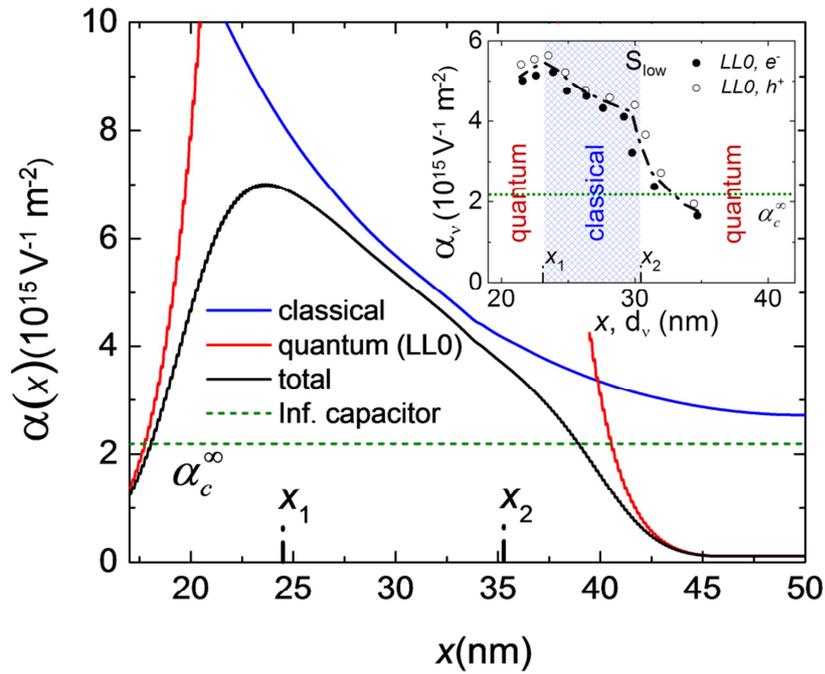

**Fig. 3**. **Capacitance profile across the nanoconstrictions of type $S_{low}$ for *LL0*.** Calculated classical (blue curve), quantum (red curve) and total effective capacitance (black curve) profiles across a graphene nanoconstriction with $W = 100$ nm. The constant dashed green line represents $\alpha_c^\infty$. The inset shows the experimental data for sample type $S_{low}$ (Fig. 2b), separating zones where classical (blue) or quantum (red) contributions dominate according to the prediction shown in the main figure. Dash-dotted lines represent a guide to the eye.

# Supplementary Information

# Gate electrostatics and quantum capacitance in ballistic graphene devices

José M. Caridad[1†], Stephen R. Power[2,3,4], Artsem A. Shylau[1], Lene Gammelgaard[1], Antti-Pekka Jauho[1], Peter Bøggild[1]

[1]*Center for Nanostructured Graphene (CNG), Department of Physics, Technical University of Denmark, 2800 Kongens Lyngby, Denmark*

[2]*Catalan Institute of Nanoscience and Nanotechnology (ICN2), CSIC and The Barcelona Institute of Science and Technology, Campus UAB, Bellaterra, 08193 Barcelona, Spain*

[3]*Universitat Autònoma de Barcelona, 08193 Bellaterra (Cerdanyola del Vallès), Spain*

[4]*School of Physics, Trinity College Dublin, Dublin 2, Ireland*

[†]corresponding author: jcar@dtu.dk

1. **Experimental details**

*Fabrication and electrical characteristics of devices prior to define nanoconstrictions*

Devices with field-effect mobility $\mu$ ~20.000 cm$^2$/Vs (estimated mean free paths $l$~200 nm) at a temperature T = 4K are achieved by the mechanical exfoliation of graphene on hydrophobic Si/SiO$_2$ substrates. [S1,S2] We select devices with relatively low contact resistances $R_c$ below 600 Ω. These initial device parameters $(\mu, R_c)$ are extracted [S2] by first shaping, contacting and measuring the transport properties of rectangular two-terminal devices with a width $W$ of ~ 1 µm (Fig. S1a), where the obtained two point resistance $R_{2pt}$ is fitted to the equation [S3]:

$$R_{2pt} = R_c + \frac{L/W}{e\mu\sqrt{n^2 + n_{res}^2}} \quad (Eq.S1).$$

Here, $e$ is the elementary charge, $n_{res}$ is the residual carrier concentration and $n$ is the back-gate induced carrier concentration given by:

$$n = \frac{C_{ox}}{e}(V_g - V_{CNP}) \quad (Eq.\ S2),$$

where $C_{ox}$ is the gate oxide capacitance per unit area for an infinite parallel-plate capacitor (3.5×10$^{-4}$ Fm$^{-2}$ for our 100nm thick SiO$_2$) and $V_{CNP}$ is the position of the charge neutrality point (CNP). Fig S1b shows the two-point conductance of the device, given by $G = 1/[R_{2pt} - R_c]$, at two different temperatures: 4K and 300 K. The decreased conductance exhibited at higher temperatures is ascribed to a decrease in the intrinsic mobility of these samples due to the role of electron-phonon scattering [S2].

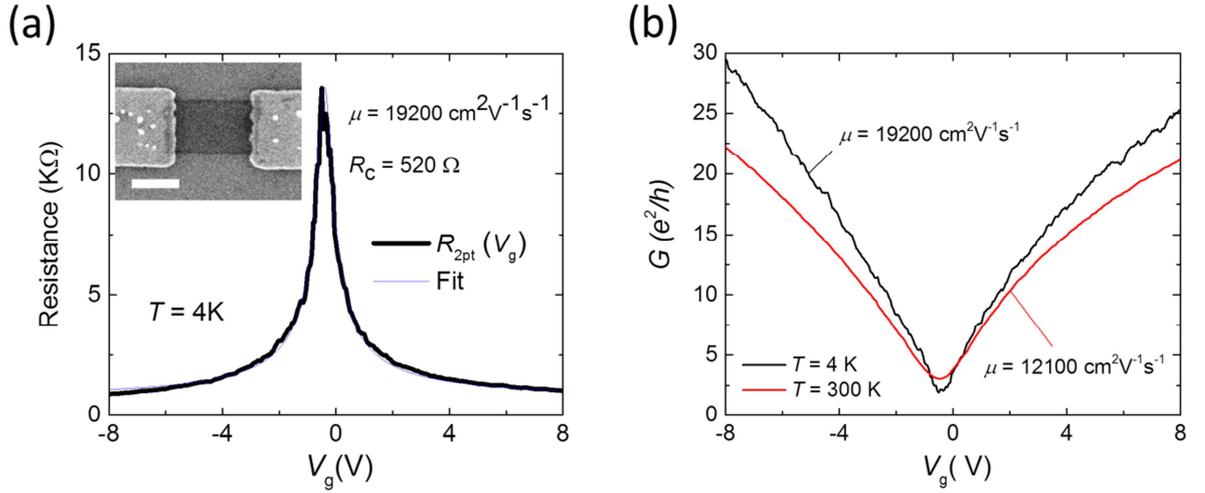

**Fig.S1.** (a) Two-point (2-pt) resistance of pre-constriction devices (graphene strips with width $W$ = 1 μm) at T = 4 K and B = 0 T. Fitting the experimental data with Eq. S1, we extract a $\mu$ = 19200 cm$^2$/Vs and a $R_C$ = 520 Ω. Inset shows a scanning electron micrograph of this device. Contacts made from Ti/Au (5/25 nm). Scale bar is 1μm. (b) Conductance of these devices at two temperatures (T = 4K and T = 300K) and B = 0 T.

Furthermore, we measure the magnetoconductance $G(V_g, B)$ of these wide graphene devices for different magnetic fields (Fig. S2a). First, this is done as an alternative method to extract the value of contact resistance in the device $R_c$. Then, from this data (Fig. 2a), we can also extract the effective capacitance of these large (and diffusive) devices following the procedure described in the main text. Fig.2b shows how this effective capacitance is nearly homogeneous, close to the values of the infinite parallel-plate capacitor and exhibits a small measured modulation below $\alpha_{tot}(x)^{(max)} / \alpha_c^\infty < 1.2$ for both *LL0* and *LL1*. These trends are expected from an electrostatic point of view for samples with $W/b \gg 1$ and are consistent with similar samples reported in literature [S4, S5].

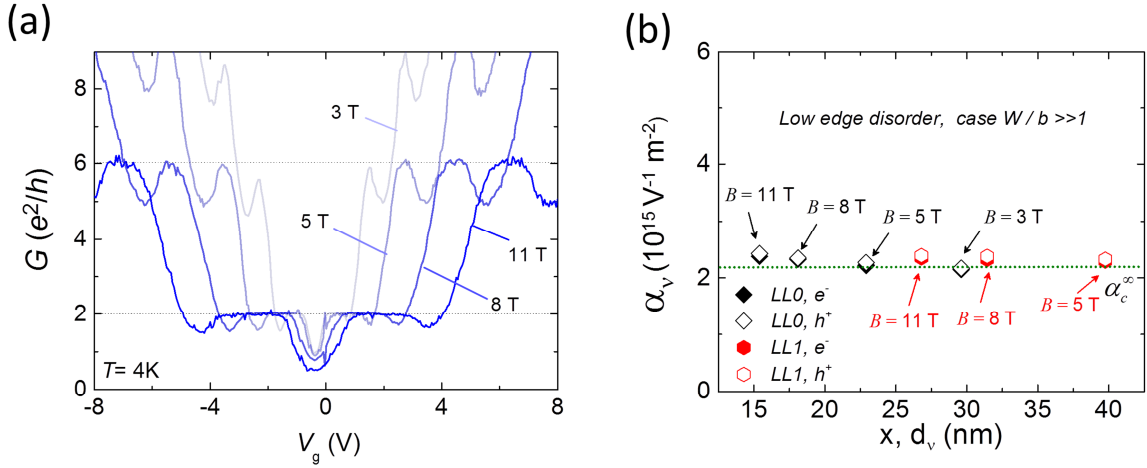

**Fig.S2.** (a) Two-point (2-pt) magnetoconductance $G(V_g, B)$ of pre-constriction devices (graphene strips of width $W$ = 1 μm) at T = 4 K and magnetic fields B = 3,5,8 and 11 T. (b) Experimental effective capacitance $\alpha_\nu$ across large devices for different Landau Levels (LLs) for electrons ($e^-$) and holes ($h^+$) at positions $x$ corresponding to the cyclotron diameter $d_\nu$.

*Fabrication of nanoconstrictions*

The subsequent definition of the nanoconstrictions is done via electron beam lithography (EBL) using polymethyl-methacrylate (PMMA) developed at -5 °C in a 1:3 IPA:H$_2$O solution. The edge quality in our constrictions is defined with two complementary etching processes: oxygen plasma ashing and reactive ion etching. [S2,S6] Devices with a higher amount of edge disorder ($S_{high}$) are defined by plasma ashing, which, despite being known to introduce instabilities and localized states in graphene nanodevices, is widely used to shape graphene nanostructures [S6]. In contrast, devices with a much lower amount of edge disorder [S2,S6] ($S_{low}$) were produced by reactive ion etching (argon 40 sccm, oxygen 5 sccm). We achieve an edge roughness below 1 nm with the latter etching procedure, as demonstrated in Ref. S2. Prior to measuring their electrical properties, we dip our devices for 18h in a pure hexamethyldisilazane (HMDS) solution to reduce the effect of environmental contaminants that may have been adsorbed on the basal plane of graphene or at the edges during the processing steps [S2,S7]. After these 18h, the devices are dipped for 5 s in acetone, 5 s in IPA and then dried with nitrogen.

Both types of nanoconstriction devices $S_{high}$ and $S_{low}$ are limited by boundary scattering [S2]. Apart from a characteristic $G \propto \Delta V_g^{1/2}$ behavior [S2], this is shown by a two-point conductance $G$ as a function of the back-gate $V_g$ not significantly changed at the two measured temperatures $T = 4$ K and T = 300K (Fig. S3).

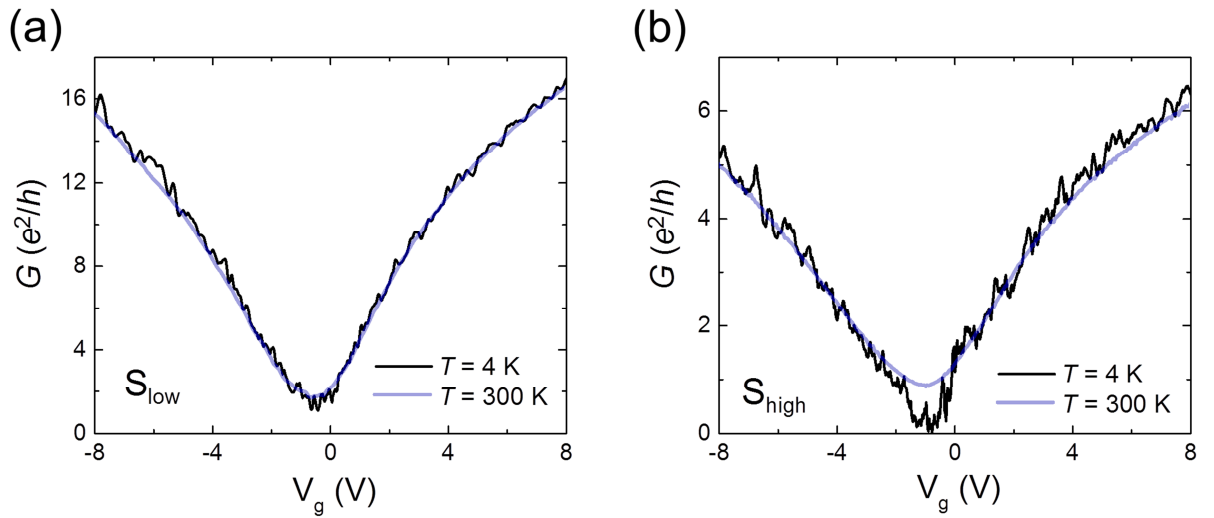

**Fig. S3.** Conductance measurements in graphene nanoconstrictions at T=4 K and T=300 K for $B$=0 T. **a**, Conductance $G$ as a function of the gate voltage $V_g$ in a graphene nanoconstriction with low edge roughness ($S_{low}$) at T = 4 K (black) and T = 300 K (blue). The extracted $R_c$ in this sample is 420 Ω. **b,** Conductance $G$ as a function of the gate voltage $V_g$ in graphene nanoconstriction with high edge roughness ($S_{high}$) at T=4K (black) and T=300K (blue). The extracted $R_c$ in this sample is 520 Ω.

2. **Validity of the semi-classical method to probe the capacitance profile in the presence of the conductance quantization suppression (CQS) phenomenon.**

The CQS phenomenon [S2,S8] is an effect occurring in ballistic and finite-size graphene devices as a consequence of the qualitative modification of the band structure of these gated systems within the quantum Hall regime (Fig. S4a) caused by fringing electrostatic field effects. As such, new, residually overlapping and counter-propagating conducting channels appear at the central part of the device with finite weight over a large portion of the device's width (Fig. S4b, local density of states LDOS). Their presence manifests itself in an increased conductance with respect to the single-particle picture, with suppressed quantization (Fig. 1d, main text) [S2]. Furthermore, as shown in Fig. S5, these features are robust with respect to temperature variations, showing small alterations in the magnetoconductance of the device that could be ascribed [S8] to the density of states (DOS) broadening by the thermal energy.

A detailed theory supporting the validity of the semi-classical method used here to probe the capacitance profile in the presence of fringing electrostatic field effects can be found in Ref. [S9]. Briefly Ref.[S9] describes how, *i)* the semi-classical approximation $n_\nu(x) = \frac{eB}{h}\nu(x)$ is valid in these systems subject to fringing electrostatic fields, where the occupation factor $\nu(x)$ varies locally across the device [S10,S11]. Furthermore, *ii)* the position of the additional counter-propagating edge channels in the system and their evolution with respect to the magnetic field is approximately given by $d_\nu = \sqrt{\frac{\nu h}{e\pi B}}$ [S9].

As such, $\nu$ can be directly obtained from the measured conductance $\nu \sim G(e^2/h)$ [S11] ] in devices showing the CQS phenomenon in the absence of major backscattering.

In our case, the experimentally observed conductance does (empirically) imply the absence of significant backscattering in nanoconstrictions with low edge roughness (otherwise counterpropagating edge channels would equilibrate [S12,S13]). This is particularly true at $G^{(max)}$, reaching values $\sim 4e^2/h$ for *LL0* at $B \leq 12$ T. Thus, it is reasonable to obtain $\nu$ at $G^{(max)}(e^2/h)$ in these systems.

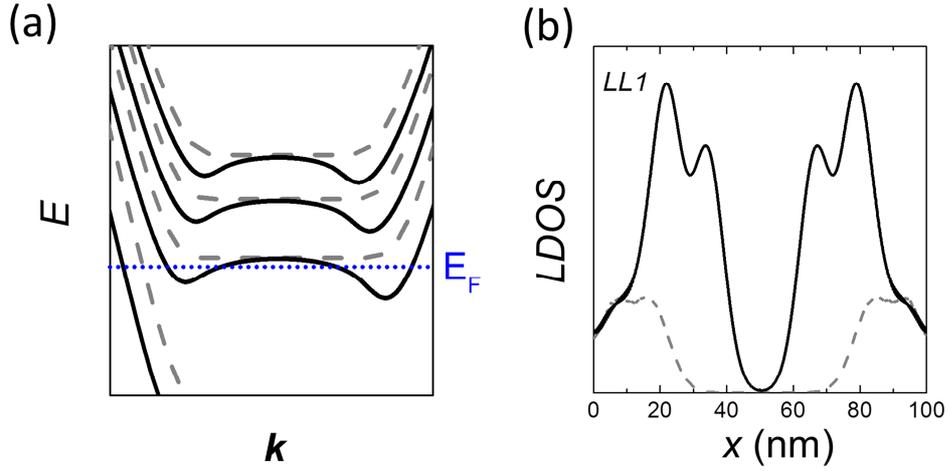

**Fig. S4. (a)** Band structure near the Fermi level $E_F$ for a ballistic graphene device in the quantum Hall regime accounting (black) and non-accounting (dashed grey) for fringe field effects. **(b)** Local density of states (LDOS) across a pristine ribbon at a gate voltage that displays a CQS peak for the *LL1*. Fringe field effects (black) introduce states in the ribbon bulk, which are not present for uniform gating (dashed gray).

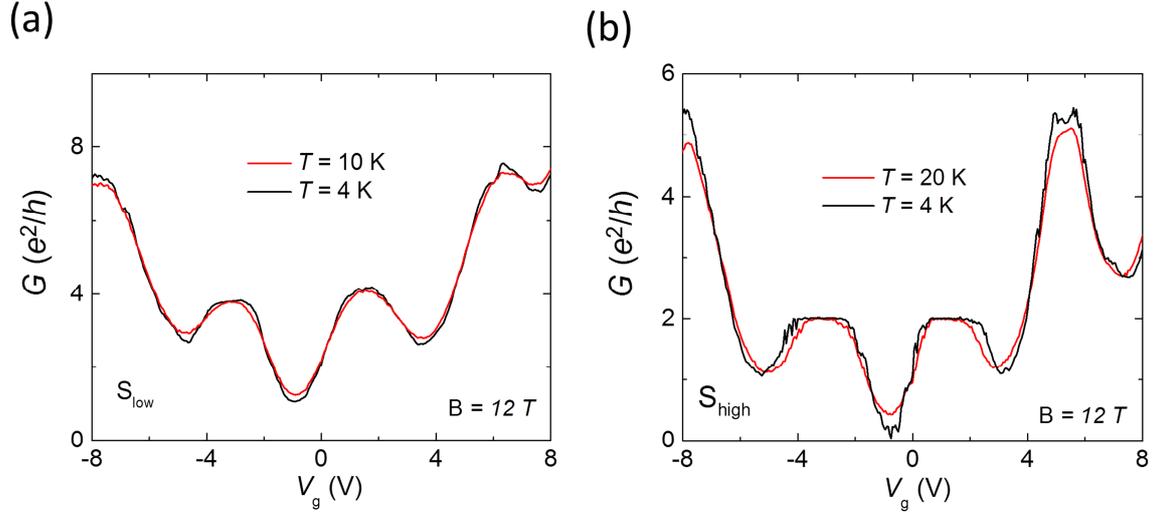

**Fig. S5**. (a) Measured two-point conductance in our nanoconstriction devices with low edge disorder ($S_{low}$) at temperatures T = 4 K and T = 10K. (b) Measured two-point conductance of our nanoconstriction devices with high edge disorder ($S_{high}$) at temperatures T = 4 K and T = 20K.

## 3. Electrostatic capacitance calculation $\alpha_c(x)$

Pristine graphene devices exhibit an inhomogeneous, classical capacitance profile $\alpha_c(x)$ diverging towards the device's edges [S4,S5,S14,S15]. $\alpha_c(x)$ can be approximately calculated considering graphene as a perfect metal [S16] and using $\alpha_c(x) = \dfrac{n_{W/2}}{V_g} \dfrac{E_z(x)}{\min(E_z(x))}$ (Fig. S6). Here, $n_{W/2}$ is carrier density at the centre of the strip (position $x = W/2$). $n_{W/2}$ is obtained by solving the Poisson equation in the device [S12,S16] using a finite-element method solver, at a distance $z = -0.5$ nm (below the graphene plane $z = 0$ nm). For our constrictions ($W = 100$ nm), its value is close to the carrier density for an infinite parallel-plate capacitor, see Fig. S6. Meanwhile, $E_z(x)/\min(E_z(x))$ is the out of plane, un-screened perpendicular electric field component in the devices [S12], normalized with respect to the minimum value $\min(E_z(x))$ which takes place at the centre of the strip, $W/2$. $E_z(x)/\min(E_z(x))$ is obtained at a distance $z = 0.5$ nm above the graphene plane.

Fig. S6 compares $\alpha_c(x)$ for our nanoconstrictions, ribbons of the same width $W = 100$ nm, and the infinite parallel-plate capacitor $\alpha_c^\infty$. Despite qualitative similarities, for a fixed position $x$ across the devices, the electrostatic capacitance profile in nanoconstrictions is larger than in nanoribbon devices of the same width.

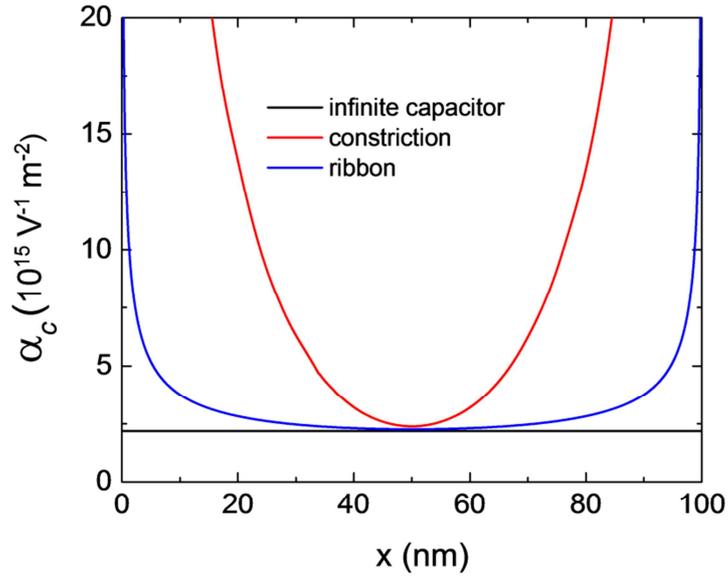

**Fig. S6**. Electrostatically calculated capacitance $\alpha_c(x)$ across three graphene devices: nanoconstriction with $W = 100$ nm (red), nanoribbon with $W = 100$ nm (blue) and infinite parallel-plate capacitor (black). For a fixed position $x$, $\alpha_c(x)$ is larger in nanoconstrictions than in nanorribbons with the same width.

4. **Quantum capacitance calculation $\alpha_q(x)$ across graphene devices (*LL0* and *LL1*)**

We calculate the quantum capacitance contribution $\alpha_q(x)$ across ballistic gated graphene devices from their corresponding LDOS as [S17,S18] $\alpha_q(x) = e|LDOS|$ (Fig. S7). In the present study, LDOS of finite-size and disorder-free graphene devices at $B = 10$T are calculated at the two gate voltages corresponding to $G^{(max)}$ for *LL0* and *LL1*, respectively. Such LDOS calculations are undertaken by using a tight-binding Hamiltonian with a corresponding $U(x) = -\hbar v_F \sqrt{\pi n(x)}$ obtained from the electrostatically capacitance profile $n(x) = \alpha_c(x) V_g$. In other words, we calculate the LDOS of the realistic system, including electrostatic screening effects of the gate potential. Further details about these numerical calculations can be found in Ref. [S2].

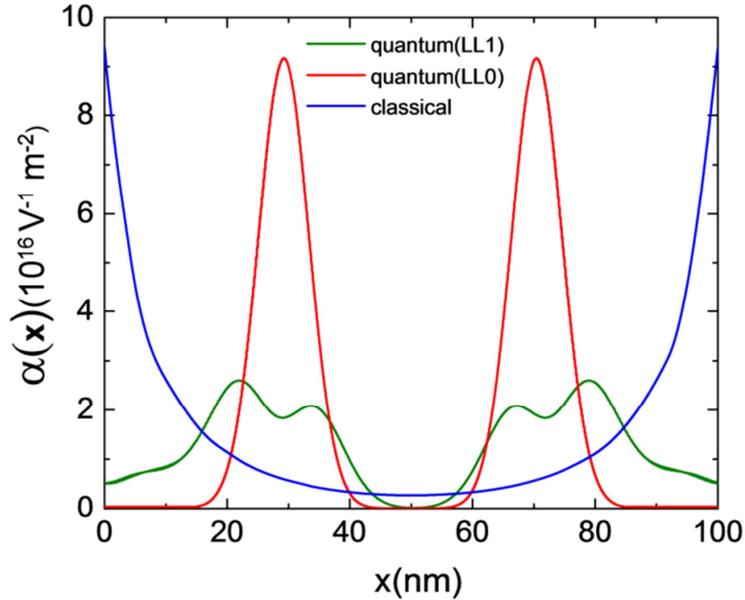

**Fig. S7. Quantum capacitance profile for *LL0* (red) and *LL1* (green) across a ballistic graphene device.** The classical capacitance profile of our nanoconstrictions (blue) is shown for comparison.